# Online placement test based on Item Response Theory and IMS Global standards

Farid Merrouch[1], Meriem Hnida[1], Mohammed Khalidi Idrissi[1] and Samir Bennani[1]

[1] RIME TEAM-Networking, Modeling and e-Learning Team- LRIE Laboratory- Research in Computer Science and Education Laboratory- Mohammadia School of Engineers (EMI) - Mohammed V University Agdal-
BP. 765 Ibn Sina Av., Agdal, Rabat, Morocco

**Abstract**

This paper aims to present an online placement test. It is based on the Item Response Theory to provide relevant estimates of learner competences. The proposed test is the entry point of our e-Learning system. It gathers the learner response to a set of questions and uses a specific developed algorithm to estimate its level. This algorithm identifies learning gaps, which allows tutors to conceive sequence of courses and remediation adapted to each case of learner, in order to achieve a competence.

***Keywords:*** *Online placement test, Item Response Theory, e-Learning Adaptive System, Learner Ability, Competence Modeling, IMS Global Standards.*

## 1. Introduction

E-Learning has always benefited from New Technologies of Information and Communication (NTIC) to continuously improve teaching and learning methods, and to refine monitoring and assessment techniques. These technologies enhance content and presentation of educational resources, make communication more efficient between individuals, and allow learners to get adapted teaching taking advantage of a feedback and support throughout a learning process. Even so, these technologies still face major problems when it comes to multiple learner profiles.

Several researchers have examined the learner profiles and proposed various adaptive e-Learning systems. the proposed systems use one of these approaches to find a solution to heterogeneous learners: (1) Modeling learner profiles in relation to educational objectives [5][15], (2) Designing learning contents adapted to learner style [1][6][13][21],(3)Creating learning paths adapted to learner achievements and progress [2][11][22][14] .

Our study goes in line with the third approach, our goal is to adapt the learning path based on the identification of learning gaps.

The main contributions of this paper are: (1) to propose a guide to an online placement test based on the Item Response Theory, (2) to estimate learner ability and identify learning gaps, (3) to suggest a competency modeling technique.

The organization of this paper is as follows: section 2 describes our research approach. Section 3 explains how Item Response Theory is used to estimate learner ability. Section 4 shows the design of our placement test and a competency modeling technique. Finally, we give in section 5, an example of an online placement test and demonstrate how learner competence is estimated.

## 2. Research approach

Considering the uniqueness of learners helps them learn and get actively involved in their learning process. However, it supposes a constant support and a regular follow-up in order to maintain a good relationship between learners and teachers in an online environment. For Chrysafiadi et al. [15], an adaptive e-Learning system should handle various learning paths adapted to everyone. It should also monitor learner's activities and deduce its needs and preferences. Yang et al. [10] has pointed out that an adaptive e-learning system has to take into account the uniqueness of learners and suggests a new student modeling technique. This technique brings a variety of information together structured as follows: (1) Static information identified in the beginning of a course. (2) Dynamic information extracted from interaction of learner and environment and (3) Contextual information as domain knowledge, or context.

Our target in this paper is to introduce an initializing technique of learner profile and learning path by a placement test. This test is based on the Item Response Theory and can be used in the beginning of a course, or as a formative assessment to update the learner profile and adjust its learning path.

## 3. Item Response Theory

Item Response Theory (IRT) is a popular measurement theory in the areas of psychology, education, sociology, linguistics, teaching, management, etc. [7]. For example, in education, IRT is used in assessment in order to (1) determine the most relevant questions to ask a learner, based on its level (2) estimate the learner ability (3) and decide whether the estimates are precise enough to stop the test. According to IRT, the learner responses during an assessment are considered as a stochastic process in which the probability of giving a correct answer or not, depends on many factors, like previous asked questions and answers [3]. It depends also on some cognitive, emotional and social components [4].

The theory aims to expose the learner to a virtual examiner able to choose adequate questions to ask. So we have used the Item Response Theory within a placement test to estimate learner ability.

3.1 IRT: Measurement Models

A model is a representation of the phenomenon to be studied, and a measure assigns numbers to objects according to specific rules or characteristics. IRT provides a unique measure of a latent trait based on a statistical model. In IRT, a model is mainly based on various estimators, related to particular individuals and conditions under which the test is done [16].

In IRT, different estimators of a person capacity can be used. According to [3] the most relevant ones are: Maximum Likelihood estimator, Bayesian estimator, Maximum a posteriori Probability (MAP). In our case, we have retained the Maximum Likelihood estimator, as we find that the error level is slightly lower than its competitors [3][8].

3.2 The probability of answering correctly a question

In a test, a learner has a probability P ($\theta$) to answer correctly an item. In education, an item refers to a question.
Once calculated, this probability can be used to give tailored questions thereafter, which fit the level of the person taking the test. The following equation shows how to calculate the probability of giving the right answer to an item. [9].

$$P(\theta) = \frac{1}{1 + e^{-a(\theta - b)}} \quad (1)$$

- $e$ : 2.718.
- $b$: difficulty parameter associated to each question.
- $a$: discrimination parameter associated to each question.

The relationship between the probability of correctly answering an item P($\theta$) and the estimated ability of a learner ($\theta$) is expressed by a function called Item Information Function and plotted with the Item Characteristic Curve. The following figure (Fig1) gives an example of an Item Characteristic Curve.

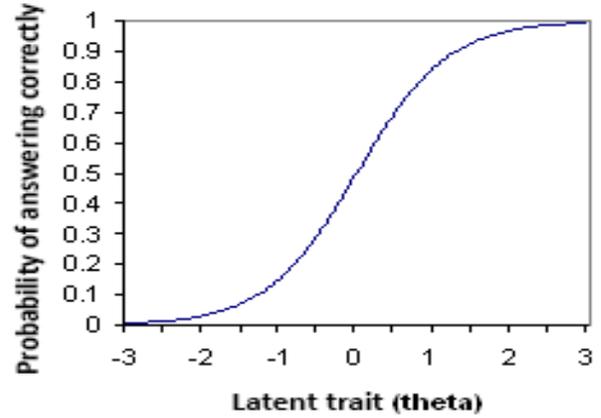

Fig1. Example of Item Characteristic Curve [9]

3.3 IRT for assessing learner competences

In a test, the result of a learner can be explained based on its knowledge but also its competence. A competence is an unobservable trait. However, we propose to give to competence a numeric value, which represents the skill level.

In IRT, a latent trait (unobservable one), is expressed by the Greek letter theta $\theta$. The following equation [9] shows how to calculate an ability or latent trait in IRT.

$$\theta_{s+1} = \theta_s + \frac{\sum_{i=1}^{N} -a_i [u_i - P_i(\theta_s)]}{\sum_{i=1}^{N} a_i^2 P_i(\theta_s) * Q_i(\theta_s)} \quad (2)$$

- $\theta_s$: Learner ability within iteration S, the value of $\theta$ is theoretically between $-\infty$ and $+\infty$ but in reality is limited between -3 and 3.
- $i$: Current asked question.
- $N$: Number of questions.
- $U_i$: The learner response to question i, ($u_i = 1$) for a correct answer and ($u_i = 0$) for a wrong one.
- $P_i(\theta_s)$: The probability to give a correct answer to question i in iteration s.
- $Q_i(\theta_s)$: The probability to give an incorrect answer to question i in iteration s.

In the next iterations: estimating a competence $\theta_{s+1}$ takes as a parameter $\theta_s$ calculated in the previous step. After each iteration, the estimation gets more precise until we obtain a stable competence value and a low error value.
To calculate the standard error we used the following equation[9]:

$$E = \frac{1}{\sqrt{\sum_{i=1}^{N} a_i^2 \, P_i(\theta_s) \, Q_i(\theta_s)}} \quad (3)$$

## 4. Design of the Placement Test

### 4.1 Modeling the learner competence to assess

We propose an online placement test that focuses on learner competence because we consider the competence as an important trait of an individual. So, the assessment of a competence is based on knowledge combined with learner characteristics which defines the competence.

To define a competence to assess, we propose a model combining Paquette's [20] and Elena's [4] modeling techniques. In what follows, we present each of these modeling techniques, and then we introduce our approach of competence modeling.

Paquette [20] defines a competence as an ability that uses a set of knowledge to achieve a learning goal [19]. An ability has a performance and reflects an action like "apply", "synthesize", "evaluate", or "memorize" a set of knowledge.

For Paquette [20] knowledge can be a concept, fact, principle or procedure.
A learning performance reflects context, complexity, autonomy, scope and frequency for a learning situation.
A performance best describes a competence. For example a context of performance shows if it's a familiar or unfamiliar (new) situation for a learner. The complexity determines how difficult a task is. Autonomy indicates whether learner needs help and assistance or not. Finally, a scope determine if the learner has to deal totally or partially with a task.
The following table (Table 1) shows the competence modeling according to Paquette [20].

**Table 1: Paquette's competence modeling technique [20]**

| Ability | Knowledge | Performance |
|---|---|---|
| Apply | Concept | Scope |
| Synthesize | Principle | Autonomy |
| Evaluate | Procedure | Situation/Context |
| Memorize | Fact | Frequency |

Elena et al [4] use UML (Unified Modeling Language) to model a competence. According to [4]:
- A competence may need the acquisition of other competences as prerequisites.
- A competence includes one or several competences called "Competency Element Definition".
- A competence can express an attitude as "Attitude Definition" or knowledge as "Knowledge Element Definition" or skills as "Skill Definition".
- A Skill requires a set of knowledge.

The following figure (Fig. 2) summarizes the competence modeling technique proposed by Elena [4].

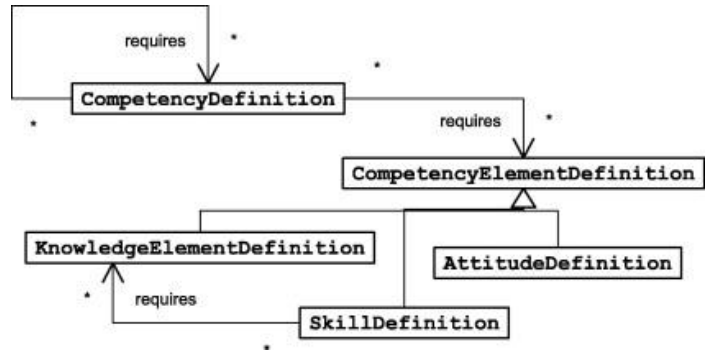

Fig2. UML competency modeling technique [4]

To obtain a complete competency definition model, we consider that the two techniques described before are complement each other. We ensure that Paquette's modeling technique [20] will help us clearly define and assess a competence and Elena et al. [4] technique will allow us organize relationship between competencies. We propose the model shown in (Fig3) to define a competence to assess in an e-Learning system. An example of use will be shown later.

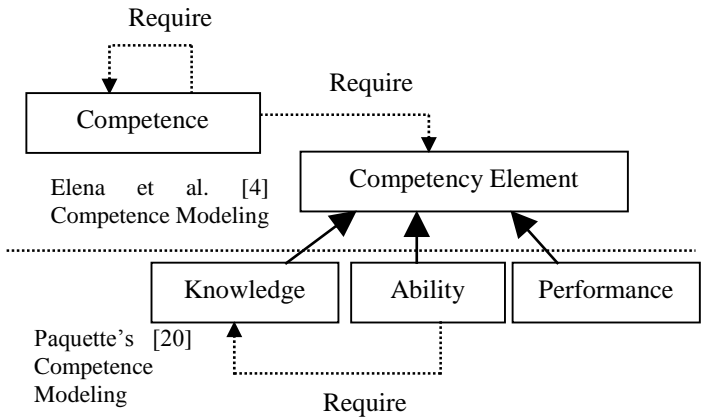

Fig3. A proposed model to define competence to assess

### 4.2 Designing assessment

The first step of designing an assessment is about gathering, in a same component, information about learner and competence to assess. The challenge was to integrate the online placement test into all available e-Learning platforms, for this purpose we have used IMS Global [12] standards as bellow:

- IMS-QTI (IMS Question & Test Interoperability) to design questions, each question definition includes: an identifier, question body (text and images), choices to present to learner, the correct answer, and scale.

- IMS-LIP (Learner Information Package) to create leaner profile, it includes information as: identification, affiliation, activity, interest, etc.

- IMS-RDCEO (Reusable Definition of Competency or Educational Objective) to implement the proposed model of competence to assess (Fig2)[18].

The second step is about implementing the Item Response Theory to assess a competence using mathematical equations explained in section 3. So, we propose an algorithm to estimate leaner ability based on IMS Global [12] standards and Item Response Theory.

The following figure (Fig 4), synthesizes the proposed designing assessment.

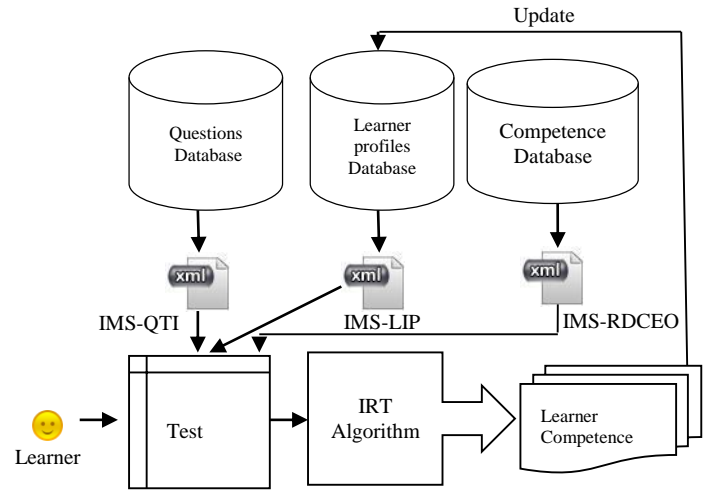

Fig4. Designing the online placement test

To assess a competence, the online test uses several informations stored in XML databases in an algorithm (explained in section 4.3). The test operates in the following way:

In the beginning, the application retrieves Items (Questions), learner profile and competency definition to construct the test.

A learner takes n questions test multiple choices with m choices for each question. The number (n) of questions is defined according to the competence to assess, and for each competence we limit a number of choices to (m). That's make of the online placement test different for one learner to another.

At the end of the assessment, the algorithm use the Item Response Theory to estimate learner ability which will be specified in his profile.

### 4.3 A proposed algorithm to assess a competence

Our proposed IRT algorithm works as follows:

- The algorithm extracts the definition of competence to assess from the e-Learning platform, filled in an XML file and based on IMS Global [12] standards.
- Once the competence to assess is available, the algorithm searches in the database of questions, all related questions.
  If the number of related questions exceeds the number of required questions, the algorithm uses the most relevant ones according to an indicator of importance.
- During assessment, the algorithm collects learner response for each question, it compares the given response with the required one, assigns 1 to correct response and 0 to an incorrect one.

- At the end, the algorithm estimate the learner ability using the equation (2) (a demonstration will be provided later)
  Then, the algorithm calculates standard errors using the equation (3).
- The algorithm follows a stochastic process: the same procedure is repeated based on previous steps for n iteration until the algorithm encounters this condition: the competence estimation gets stable after n iteration.
- Finally, the algorithm fills the competence value in the learner profile and displays the result to learner.

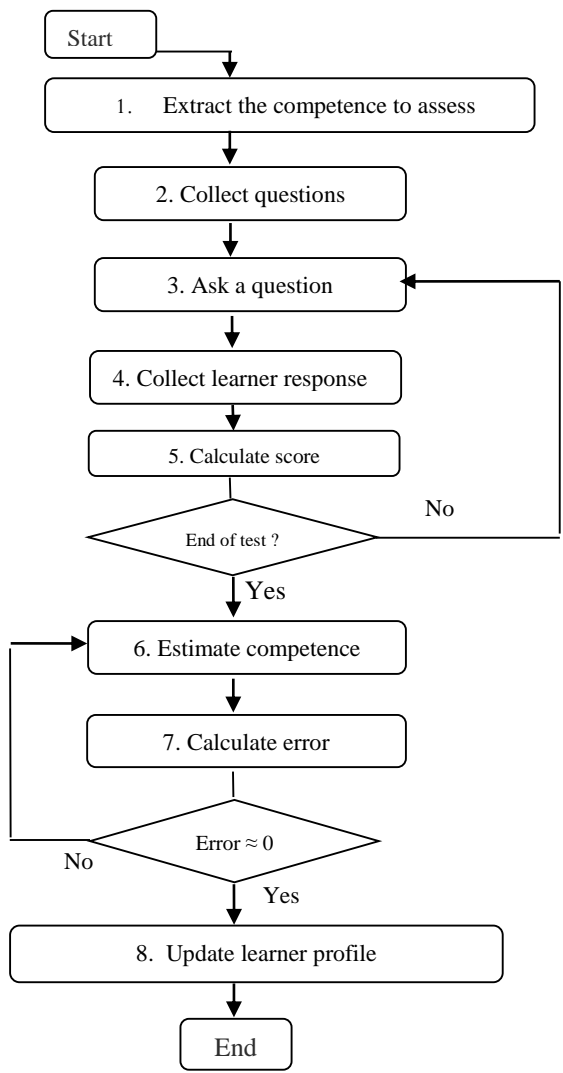

Fig5. Proposed IRT algorithm

## 5. A placement test example

### 5.1 Example of competence to assess

To give an example of a placement test we propose 20 questions related to a competence in "SQL Structured Query Language".
To acquire a competence in SQL we assume that:
- A competence in SQL requires a prior competence in relational algebra.
- To acquire a competence in SQL, the learner must apply SQL to create data structures, manipulate data and retrieve data from a schema.
- As competence in SQL needs the mastery of creating data, manipulating and retrieving data from a schema, each of these is considered as element of competence with his own knowledge, and desired performance and ability to acquire.

The figure (Figure 6) illustrates a model of SQL competence to assess based on the model in (Figure 3)

### 5.2 Explanation of the proposed IRT algorithm

We suppose that a learner answered incorrectly the questions 1,4,7,8,15,16,18,19 (value 0 of Ui) and answered correctly to the other questions (value 1 of Ui).
The first column of the table (Table 2) corresponds to question number, called also item or "i".
The response of the learner "Ui" to question "i" takes a value of 1 if the response given by learner is a correct answer and 0 if it is not.
In table 2 the blue lines indicates incorrect answers.

The algorithm uses the formulas (1) and (2) mentioned above including:
- The difficulty parameter, denoted by b, increases over the questions with a value of 0.1. Questions become more and more difficult. So, we found that:
- Pi: the probability to give a correct answer decreases over the questions.
- Qi: the probability to give an incorrect answer increases over the questions.

- **Step 1 : calculate the probability of answering correctly a question i**

*For question number 1:* the difficulty parameter is 0.1.

Using the equation (1) we obtain:
$$P_1(1) = \frac{1}{1+ e^{-1(1-0.1)}} = 0.7109303258$$

The learner has 71% chance of giving the right answer.
We obtain also:
$Q_1(1) = 1 - P_1(1) = 1 - 0.7109303258129135$

$Q_1(1) = 0.2890696741870865$.

The learner has 29% chance of giving the wrong answer.

*For question number 2:*
The difficulty parameter is 0.2. Using equation (1), we obtain:

$$P_2(1) = \frac{1}{1+e^{-1(1-0.2)}} = 0.6899567375$$

The learner has 69% chance of giving the right answer.

*And $Q_2(1) = 1 - P_2(1) = 1 - 0.6899567375912588$*
$Q_2(1) = 0.31004326240874125$

The learner has 31% chance of giving the wrong answer.
We repeat the same operation for the next n question.

**Step 2: Estimate the competence used to answer each question**

For each question we calculate **$a_i[u_i - P_i(\theta s)*]$** and **$a_i^2 P_i(\theta s) * Q_i(\theta s)$** which are numerator and denominator of equation (2), Ui is the response of learner to question i.

**For question number 1:**
- The numerator:
$a_1[u_1 - P_1(1)] = 1(0 - 0.7109303258129135)$
$a_1[u_1 - P_1(1)] = -0.7109303258129135$

- The denominator :
$a_1{}^2 P_1(\theta s) * Q_1(\theta s) =$
$1*(0.7109303258129135)*(0.2890696741870865)$
$= 0.20550839765245815$

**For question number 2:**
- The numerator:
$a_2[u_2 - P_2(1)] = 1(1 - 0.6899567375912588)$
$a_2[u_2 - P_2(1)] = 0.31004326240874125$

- The denominator:
$a_2{}^2 P_2(1) * Q_2(1) =$
$1*(0.6899567375912588)*(0.31004326240874125)$
$= 0.21391643784368566$

We repeat the same operation for the next n questions.

The following table records the results of the algorithm (table.2) in the first iteration:

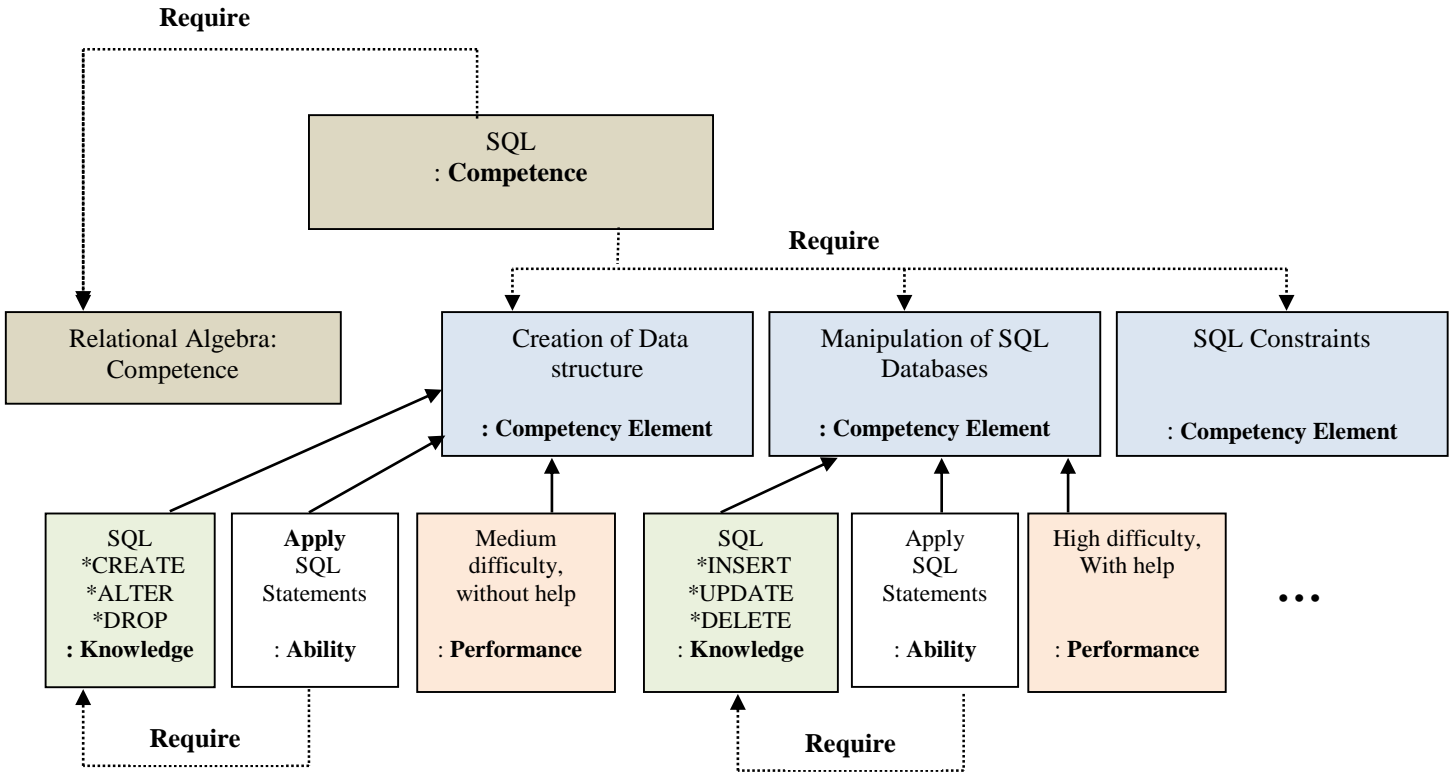

Fig6. Example of competence modeling

**Table 2: Result of the algorithm, iteration 1**

| i | $U_i$ | b | $P_i$ | $Q_i$ | Num | Denom |
|---|---|---|---|---|---|---|
| 1 | 0 | 0.1 | 0,7109 | 0,2891 | -0,7109 | 0,2055 |
| 2 | 1 | 0,2 | 0,6900 | 0,3100 | 0,3100 | 0,2139 |
| 3 | 1 | 0, 3 | 0,6682 | 0,3318 | 0,3318 | 0,2217 |
| 4 | 0 | 0,4 | 0,6456 | 0,3544 | -0,6456 | 0,2288 |
| 5 | 1 | 0,5 | 0,6224 | 0,3776 | 0,3776 | 0,2350 |
| 6 | 1 | 0,6 | 0,5987 | 0,4013 | 0,4013 | 0,2403 |
| 7 | 0 | 0,7 | 0,5744 | 0,4256 | -0,5744 | 0,2445 |
| 8 | 0 | 0,8 | 0,5498 | 0,4502 | -0,5498 | 0,2475 |
| 9 | 1 | 0,9 | 0,5250 | 0,4750 | 0,4750 | 0,2494 |
| 10 | 1 | 1,0 | 0,5000 | 0,5000 | 0,5000 | 0,2500 |
| 11 | 1 | 1,1 | 0,4750 | 0,5250 | 0,5250 | 0,2494 |
| 12 | 1 | 1,2 | 0,4502 | 0,5498 | 0,5498 | 0,2475 |
| 13 | 1 | 1,3 | 0,4256 | 0,5744 | 0,5744 | 0,2445 |
| 14 | 1 | 1,4 | 0,4013 | 0,5987 | 0,5987 | 0,2403 |
| 15 | 0 | 1,5 | 0,3776 | 0,6224 | -0,3776 | 0,2350 |
| 16 | 0 | 1,6 | 0,3544 | 0,6456 | -0,3544 | 0,2288 |
| 17 | 1 | 1,7 | 0,3318 | 0,6682 | 0,6682 | 0,2217 |
| 18 | 0 | 1,8 | 0,3100 | 0,6900 | -0,3100 | 0,2139 |
| 19 | 0 | 1,9 | 0,2891 | 0,7109 | -0,2891 | 0,2055 |
| 20 | 1 | 2,0 | 0,2690 | 0,7310 | 0,7310 | 0,1966 |
| | | | | SUM | 2,23104 | 4,61973 |

**Table 3: result of the algorithm, iteration 2**

| i | $U_i$ | b | $P_i$ | $Q_i$ | Numerator | Denominator |
|---|---|---|---|---|---|---|
| 1 | 0 | 0.1 | 0,7994 | 0,2006 | -0,7994 | 0,1603 |
| 2 | 1 | 0,2 | 0,7829 | 0,2171 | 0,2171 | 0,1700 |
| 3 | 1 | 0,3 | 0,7655 | 0,2345 | 0,2345 | 0,1795 |
| 4 | 0 | 0,4 | 0,7470 | 0,2530 | -0,7470 | 0,1890 |
| 5 | 1 | 0,5 | 0,7277 | 0,2723 | 0,2723 | 0,1982 |
| 6 | 1 | 0,6 | 0,7074 | 0,2926 | 0,2926 | 0,2070 |
| 7 | 0 | 0,7 | 0,6863 | 0,3137 | -0,6863 | 0,2153 |
| 8 | 0 | 0,8 | 0,6644 | 0,3356 | -0,6644 | 0,2230 |
| 9 | 1 | 0,9 | 0,6417 | 0,3583 | 0,3583 | 0,2299 |
| 10 | 1 | 1,0 | 0,6184 | 0,3816 | 0,3816 | 0,2360 |
| 11 | 1 | 1,1 | 0,5946 | 0,4054 | 0,4054 | 0,2411 |
| 12 | 1 | 1,2 | 0,5703 | 0,4297 | 0,4297 | 0,2451 |
| 13 | 1 | 1,3 | 0,5456 | 0,4544 | 0,4544 | 0,2479 |
| 14 | 1 | 1,4 | 0,5207 | 0,4793 | 0,4793 | 0,2496 |
| 15 | 0 | 1,5 | 0,4957 | 0,5043 | -0,4957 | 0,2500 |
| 16 | 0 | 1,6 | 0,4708 | 0,5292 | -0,4708 | 0,2491 |
| 17 | 1 | 1,7 | 0,4460 | 0,5540 | 0,5540 | 0,2471 |
| 18 | 0 | 1,8 | 0,4214 | 0,5786 | -0,4214 | 0,2438 |
| 19 | 0 | 1,9 | 0,3972 | 0,6028 | -0,3972 | 0,2394 |
| 20 | 1 | 2,0 | 0,3736 | 0,6264 | 0,6264 | 0,2340 |
| | | | | SUM | 0.0243 | 4,5452 |

In order to calculate the learner competence. We give the algorithm an initial value of $\theta_0$ equal to 1 and we use the equation (2) as bellow:

$$\theta_{s+1} = \theta_s + \frac{\sum_{i=1}^{N} -a_i [u_i - P_i(\theta_s)]}{\sum_{i=1}^{N} a_i^2 P_i(\theta_s) * Q_i(\theta_s)}$$

$$\theta_1 = \theta_0 + \frac{2,23104}{4,61973}$$

$$\theta_1 = 1 + \frac{2,23104}{4,61973}$$

$\theta_1 = 1.4829370716345234$

The learner competence is **1.483**.

The following table records the results of the algorithm (Fig5) in the second iteration:

In iteration 2, we repeat the same procedure. We use the estimation of learner competence $\theta_1=1,483$ (Obtained in iteration 1): if we need to calculate $\theta_2$ we need $\theta_1$ as parameter.

- **Step 1 : calculate the probability of answering correctly a question i (Second iteration)**

*For question number 1:*

The difficulty parameter is 0.1.

$$P_1(1, 4829) = \frac{1}{1 + e^{-1(1,4829-0.1)}} = 0.7994393031$$

Learner has 80% chance of giving the right answer to question 1.

$Q_1(1, 4829) = 1 - P_1(1, 4829) = 1 - 0.7994393031976526$
$Q_1(1, 4829) = 0.20056069680234745$

Learner has 20% chance of giving the wrong answer to question 1.

*For question number 2:*
The difficulty parameter is 0.2. Using equation (1), we obtain:

$$P_2(1,4829) = \frac{1}{1+ e^{-1(1-0.2)}} = 0.7829267099$$

The learner has 78% chance of giving the right answer to question 2.

$Q_2(1, 4829) = 1 - P_2(1, 4829) = 1 - 0.7829267099347529$
$Q_2(1, 4829) = 0.21707329006524712$

The learner has 22% chance of giving the wrong answer to question 2.
We repeat the same operation for the next n question.

**Step 2: Estimate the competence used to answer each question (second iteration)**
For each question we calculate $a_i[u_i - P_i(\theta s)*]$ and $a_i^2 P_i(\theta s) * Q_i(\theta s)$ which are numerator and denominator of equation (2), $U_i$ is the response of learner to question i.

*For question number 2:*
- The numerator:
$a_1[u_1 - P_1(1, 4829)] = 1(0 - 0.7994393031976526)$
$a_1[u_1 - P_1(1, 4829)] = -0.7994393031976526$

- The denominator:
$a_1^2 P_1(\theta s) * Q_1(\theta s) =$
$1*(0.7994393031976526)*(0.20056069680234745)$
$= 0.1603361037005043$

*For question number 2:*
- The numerator:
$a_2[u_2 - P_2(1, 4829)] = 1(1 - 0.7829267099347529)$
$a_2[u_2 - P_2(1, 4829)] = -0.21707329006524712$

- The denominator:
$a_2^2 P_2(1, 4829) * Q_2(1, 4829) =$
$1*(0.7829267099347529)*(0.21707329006524712)$
$= 0.16995247680549622$

After N iterations (Table 4) we obtain a precise estimation competence, a value between [-3, 3]. In this example, the learner competence is estimated to 1.48820, and is obtained after 19 iteration, using the equation (2).

Table 3: Result of the algorithm, iteration 19

| i | $U_i$ | b | $P_i$ | $Q_i$ | Num | Denom |
|---|---|---|---|---|---|---|
| 1 | 0 | 0.1 | 0,8003 | 0,1997 | -0,8003 | 0,1598 |
| 2 | 1 | 0,2 | 0,7838 | 0,2162 | 0,2162 | 0,1694 |
| 3 | 1 | 0,3 | 0,7664 | 0,2336 | 0,2336 | 0,1790 |
| 4 | 0 | 0,4 | 0,7480 | 0,2520 | -0,7480 | 0,1885 |
| 5 | 1 | 0,5 | 0,7287 | 0,2713 | 0,2713 | 0,1977 |
| 6 | 1 | 0,6 | 0,7085 | 0,2915 | 0,2915 | 0,2065 |
| 7 | 0 | 0,7 | 0,6874 | 0,3126 | -0,6874 | 0,2149 |
| 8 | 0 | 0,8 | 0,6656 | 0,3344 | -0,6656 | 0,2226 |
| 9 | 1 | 0,9 | 0,6429 | 0,3571 | 0,3571 | 0,2296 |
| 10 | 1 | 1,0 | 0,6197 | 0,3803 | 0,3803 | 0,2357 |
| 11 | 1 | 1,1 | 0,5958 | 0,4042 | 0,4042 | 0,2408 |
| 12 | 1 | 1,2 | 0,5715 | 0,4285 | 0,4285 | 0,2449 |
| 13 | 1 | 1,3 | 0,5469 | 0,4531 | 0,4531 | 0,2478 |
| 14 | 1 | 1,4 | 0,5220 | 0,4780 | 0,4780 | 0,2495 |
| 15 | 0 | 1,5 | 0,4971 | 0,5029 | -0,4971 | 0,2500 |
| 16 | 0 | 1,6 | 0,4721 | 0,5279 | -0,4721 | 0,2492 |
| 17 | 1 | 1,7 | 0,4473 | 0,5527 | 0,5527 | 0,2472 |
| 18 | 0 | 1,8 | 0,4227 | 0,5773 | -0,4227 | 0,2440 |
| 19 | 0 | 1,9 | 0,3985 | 0,6015 | -0,3985 | 0,2397 |
| 20 | 1 | 2,0 | 0,3748 | 0,6252 | 0,6252 | 0,2343 |
| | | | SOMME | | -1,1102 | 4,4512 |

## 6. Concluding remarks

Assessing learner competencies in the beginning of learning is interesting. It's a diagnosis of learning gaps and needs. The proposed on line placement test focuses on each case of learner and gives question according to the competence we would like to assess.

In the near future, we hope creating homogenous groups of online learners according to their competencies but also their interests.

For this, we expect to use the algorithm we developed, and incorporate the concept of Zone of Proximal Development (ZPD) a concept proposed by psychologist Lev Vygotski explained in [17]. The ZPD is the distance between what a learner can do or learn alone and what a learner should do with other learners in a group. A useful concept, which will help us separate the parts in which the learner must receive individualized learning and those in which it should be part of a group.

## 7. Conclusion

The Item Response Theory is a strong mathematical theory, it has shown its effectiveness in adaptive testing. The estimation of individuals' ability in education, sports, linguistics, etc. are precise but not enough: the online placement test is only an opportunity for learner to show his competence: that's means if the learner answers by guessing, or doesn't invest his true competence into the test the obtained estimations won't be reliable.

Among the advantages of this theory is that it provides an estimation of competence whatever the questions are or the level of the learner is. The estimation varies depending on the competence and commitment of the learner taking the test.

**F. Merrouch**, Engineer degree in Computer Science in 1999; Extended Higher Studies Diploma in software engineering in 2006 from Mohammadia School of Engineers (EMI); PhD Student in Computer Science; Areas of interest: Model driven engineering, patterns, semantically web and eLearning content engineering.

**M. Hnida,** Engineer degree in Computer Science in 2012 From National School of Applied Sciences (ENSA); PhD Student in Computer Science at Mohammadia School of Engineers (EMI), Areas of interest: Adaptive e-Learning systems, Competency-Based Approach, Individualized learning path.

**M. Khalidi Idrissi,** Doctorate degree in Computer Science in 1986, PhD in Computer Science in 2009; Former Assistant chief of the Computer Science Department at the Mohammadia School of Engineers (EMI); Pedagogical Tutor of the Computer Science areas at the Mohammadia School of

Engineers (EMI) Professor at the Computer Science Department-EMI; 14 recent publications papers between 2011 and 2014; Ongoing research interests: Software Engineering, Information System, Modeling, MDA, ontology, SOA, Web services, eLearning content engineering, tutoring, assessment and tracking,

**S. Bennani,** Engineer degree in Computer Science in 1982; PhD in Computer Science in 2005; Former chief of the Computer Science Department at the Mohammadia School of Engineers (EMI); Professor at the Computer Science Department-EMI; 14 recent publications papers between 2011 and 2014; Ongoing research interests: SI, Modeling in Software Engineering, Information System, eLearning content engineering, tutoring, assessment and tracking.